\documentclass[11pt]{article}
\usepackage{graphics}

\begin{document}
\title{A counterexample against the Vlasov equation}
\author{C. Y. Chen\\
Department of Physics, Beijing University of Aeronautics\\
and Astronautics, Beijing, 100083, P.R.China
\thanks{Email: cychen@buaa.edu.cn}}
\maketitle

\vskip 50pt

\begin{abstract}
A simple counterexample against the Vlasov equation is put
forward, in which a magnetized plasma is perturbed by an
electromagnetic standing wave.
\end{abstract}

In today's physics, the Vlasov equation serves as the very
foundation for a great number of formulations and calculations in
fields ranging from star physics to controlled fusion.

The equation states that the distribution function $f(t,{\bf
r},{\bf v})$ of a collection of moving charged particles (called a
plasma in the usual physics language) satisfies\cite{krall,ich}
\begin{equation} \label{vla}
\frac{\partial f}{\partial t}+
 {\bf v}\cdot \frac{\partial f}{\partial {\bf r}}+
\frac q m ({\bf E}+ {\bf v}\times {\bf B})\cdot \frac{\partial
f}{\partial {\bf v}} =0,
\end{equation}
where $({\bf E}, {\bf B})$ represents the electromagnetic field
experienced by the particles of charge $q$ and mass $m$ at the
position $\bf r$ and with the velocity $\bf v$. It is sometimes
called the collisionless Boltzmann equation in statistical
mechanics due to the fact that the equation takes almost the same
form of the ordinary Boltzmann equation with the collision term
neglected. In an alternative form, Eq. (\ref{vla}) is written
as\cite{krall,reif}
\begin{equation} \label{vla1} \left. \frac{d f[t,{\bf r}(t),{\bf
v}(t)]}{d t} \right|_{\rm path} =0,
\end{equation}
where the subscripts imply that the derivative is performed along
a particle's path. This expression can be interpreted as the
path-invariance of the distribution function in the
six-dimensional phase space spanned by the position ${\bf
r}=(x,y,z)$ and velocity ${\bf v}=(v_x,v_y,v_z)$. Recognized as a
good statistic approximation by this community, Eq. (\ref{vla}),
or (\ref{vla1}), is very often used to investigate the dynamical
behavior of dilute plasma.

Being exposed by the fact that the regular Boltzmann equation
suffers substantial difficulties\cite{chen1,chen2}, the author has
somehow been compelled to find out in what situations the Vlasov
equation can be falsified. It turns out that a number of
counterexamples against the Vlasov equation can be constructed in
plain manners. In this paper, we shall focus ourselves on one of
them, in which the six-dimensional physics is reduced to a
one-dimensional one and can be understood with conceptual clarity.

Let's start with a plasma having a strong constant magnetic field
$B_0$ along the $z$-direction. Suppose that its distribution
function $f\equiv f(t,{\bf r},{\bf v})$ is initially uniform in
the position space, and that it is initially independent of
velocity when the speeds of the concerned particles are below a
certain limit $v_0$. Namely, we have, with $v<v_0$ and at $t=0$,
\begin{equation}\label{con1}
 \frac {\partial f}{\partial x}=\frac
{\partial f}{\partial y}=\frac {\partial f}{\partial z}= \frac
{\partial f}{\partial v_x}=\frac {\partial f}{\partial v_y}=\frac
{\partial f}{\partial v_z} = 0 ;
\end{equation}
or, in another form,
\begin{equation}\label{fc}
f(t,{\bf r},{\bf v})|_{t=0}= f_c,\end{equation} where $f_c$
represents a constant. For further simplicity, we also assume that
the particles whose speeds are beyond $v_0$ occupy only a small
proportion of the plasma and can be disregarded in our
consideration.

Adopting all the initial conditions mentioned above, we see that
no matter what kind of electromagnetic field is applied, Eq.
(\ref{vla}) leads us to nothing but
\begin{equation}\label{ft} \left.\frac{\partial f}{\partial t}
\right|_{t=0} =0,\end{equation} which means that the distribution
function of the system does not change at the beginning. More than
that, Eq. (\ref{vla1}) informs us that if an observer adheres to
any moving particle, he/she finds that $f$ is equal to $f_c$
constantly, irrespective of the position and velocity, which
literally infers that the distribution function remains unchanged
at the beginning and the later time; or, in terms of macroscopic
quantities there is no density change and no current everywhere
once for all (readers may also refer to Chapter 8 of Ref. 1, in
which the linearized Vlasov equation for uniformly magnetized
plasmas yields the same conclusion).

However, physical inspection tells us a very different story:
applying many types of electromagnetic fields will change the
plasma significantly. For this to be seen in an intuitive and
heuristic manner, let the magnetized plasma be perturbed by an
electromagnetic standing wave in the form:
\begin{equation}
\label{ele_mag} E_{1y}=\displaystyle\frac{a}{k_x} \cos(\omega t)
\sin(k_x x),\quad  B_{1z}= -\frac{a}{\omega} \sin(\omega t)
\cos(k_x x),\end{equation} where $a$ is a small constant. Note
that this perturbative wave obeys Maxwell's equation
$\nabla\times{\bf E}=-{\partial {\bf B}}/{\partial t}$.

As well known in physics, while all particles in the plasma make
gyration due to the existence of $B_0$, the gyro-center of every
particle (guiding center) involves two types of relatively slow
drifts due to the existence of $E_{1y}$ and $B_{1z}$. The first
slow drift is the ${\bf E}\times {\bf B}$ drift  along the
$x$-direction:
\begin{equation}\label{EB} \bar v_x=\frac
{E_{1y}}{B_0}=\frac{a}{k_xB_0} \cos(\omega t) \sin(k_x
x),\end{equation} and the second slow drift the grad-B drift along
the $y$-direction
\begin{equation}\label{gradB}
\bar v_y=\frac {v_\perp \rho}{2B_0}\cdot \frac {\partial
B_{1z}}{\partial x}=  \frac {a k_x v_\perp \rho}{2 \omega B_0}
\sin(\omega t) \sin(k_x x). \end{equation} Equations (\ref{EB})
and (\ref{gradB}) hold when $\omega$ is much smaller than the
gyro-frequency and the perturbed fields are relatively weak.

If observing the plasma in terms of measuring velocity and speed,
two things can be found. The first is that expression (\ref{EB})
apparently represents an average velocity  along the $x$-direction
(forming a macroscopic current if the plasma is a single-component
one). The second is that there is a speed change related to every
particle due to the combined effect of the grad-B drift and the
electric field $E_{1y}$ although the change, of $a^2$-order in
magnitude, is relatively unimportant in a perturbative approach.

If the density change is of concern, something against the
standard theory will also surface. While the grad-B drift
expressed by (\ref{gradB}) is uniform along the $y$-direction,
causing no density change anywhere, the ${\bf E}\times {\bf B}$
drift expressed by (\ref{EB}) is nonuniform along the
$x$-direction and must result in a density change along the
$x$-direction. By assuming that the gyro-radii of all the
concerned particles are relatively small (much smaller than
$2\pi/k_x$) and noticing the fact that the ${\bf E}\times {\bf B}$
drift is velocity-independent, the density change can be evaluated
with ease. With help of the linearized continuity equation of
ordinary fluid, we find that
\begin{equation}\label{n1}
\frac{\partial n_1}{\partial t} =-n_0 \frac{\partial \bar
v_x}{\partial x} =- \frac {n_0 a}{B_0}\cos(\omega t) \cos(k_x
x),\end{equation} where $n_0$ is the initial density of the
plasma. Fig.{\hskip 4pt}1 demonstrates that particles in this
plasma, regardless of their velocities, converge and diverge in
the pattern revealed by (\ref{n1}).

\includegraphics*[130,430][500,640]{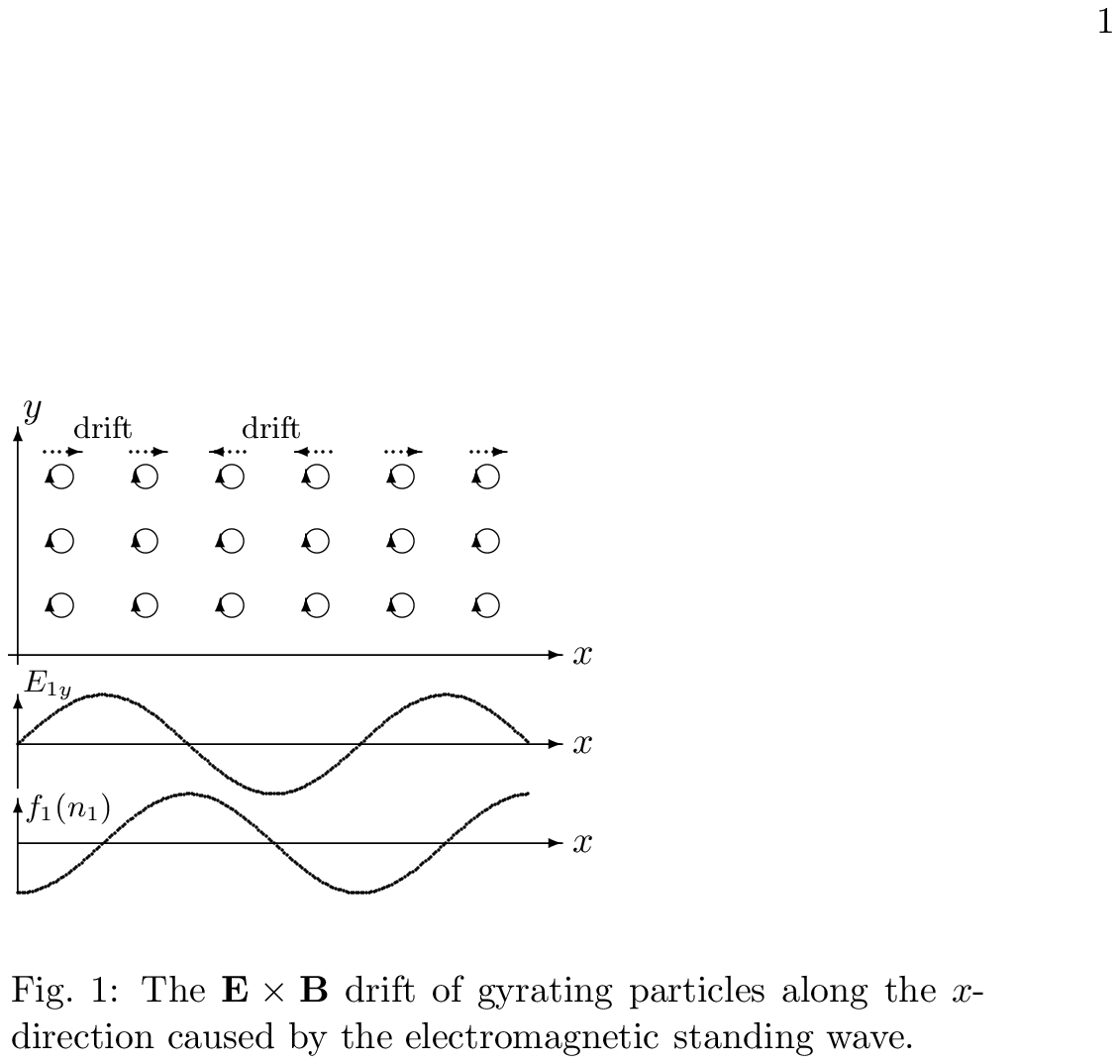}

It is at this stage intriguing and essential to look at what
happens in the phase space spanned by $x,y,z,v_x,v_y,v_z$. It
should be pointed out that this six-dimensional space is a truly
abstract and weird one, serving as a great source of all
confusions and mistakes. Investigating the change of $f$ in the
phase space challenges our conceptual and experimental wisdom
though it is surely in the domain of classical mechanics.
Referring to Fig.{\hskip 4pt}1, let's imagine that there is a
small virtual box adhering to a moving particle, represented by
$\Delta x\Delta y\Delta z$ symbolically, and imagine that we are
able to count the number of the inside particles whose velocities
are within a small range, represented by $\Delta v_x\Delta
v_y\Delta v_z$ symbolically. What will be seen by us when the box
$\Delta x\Delta y\Delta z$ drifts leftwards and rightwards in the
circumstances? At the two ending points, the density in the box
$\Delta x\Delta y\Delta z$ will, as revealed above, change from
the lowest to the highest (or vice versa) while the velocities of
the particles inside the box keep almost unchanged (noticing that
the drift velocity at the ending points are zero and the speed
changes of the particles are of second order). All this means that
$f$ in the phase volume element $\Delta x\Delta y\Delta z\Delta
v_x\Delta v_y\Delta v_z$ is not invariant along a particle's path.

Many theoretical and practical questions then arise. If the Vlasov
equation is indeed unsound, how come there exist so many
academical proofs showing otherwise? If the standard theory fails
to yield good predictions about density's changes and drift
currents, why haven't present computational simulations revealed
any discrepancy between the theory and reality? In what way can we
correctly formulate the behavior of plasma? Extensive effort has
been made by the author to answer these questions and to find out
new things\cite{chen2,chen3}, but the greatest challenge was, and
still is, to convince the mainstream of this community to have
open mind on related issues.

The author thanks professors Oliver Penrose and Keying Guan for
stimulating discussions.

\end{document}